\documentclass[12pt,superscriptaddress,showpacs]{revtex4-2}
\usepackage[T2A]{fontenc}
\usepackage[utf8]{inputenc}
\usepackage{color}
\usepackage[russian,english]{babel}
\usepackage{units}
\usepackage{amsmath}
\usepackage{graphicx}
\usepackage[bookmarks=false,
 breaklinks=false,pdfborder={0 0 0},pdfborderstyle={},backref=false,colorlinks=true]
 {hyperref}
\hypersetup{
 citecolor=blue,urlcolor=blue,linkcolor=blue}

\makeatletter

\InputIfFileExists{t2aenc.def}{}{%
  \errmessage{File `t2aenc.def' not found: Cyrillic script not supported}}

\providecommand{\tabularnewline}{\\}
\newcommand{\lyxdot}{.}

\usepackage{bm}

\makeatother

\begin{document}
\title{Bremsstrahlung on noble gases at low energies}

\author{A.I. Milstein}
\email{A.I.Milstein@inp.nsk.su}

\author{S.G. Salnikov}
\email{S.G.Salnikov@inp.nsk.su}

\affiliation{Budker Institute of Nuclear Physics of SB RAS, 630090 Novosibirsk,
Russia}
\affiliation{Novosibirsk State University, 630090 Novosibirsk, Russia}

\author{M.G. Kozlov}
\email{kozlov\_mg@pnpi.nrcki.ru}

\affiliation{Petersburg Nuclear Physics Institute of NRC “Kurchatov Institute”,
    188300 Gatchina, Russia}
\affiliation{St. Petersburg Electrotechnical University “LETI”, Prof. Popov Str. 5,
    197376 St. Petersburg, Russia}

\date{\today}

\begin{abstract}
A detailed analysis of the bremsstrahlung spectrum at nonrelativistic
electron scattering on argon and xenon is carried out.
It is shown that the approximate formulas widely used for the description of
bremsstrahlung spectra lead to predictions that significantly
differ from the exact results. In the limit when the photon frequency
tends to zero, a rigorous proof of the relationship between the spectrum of the bremsstrahlung
with a transport cross section of electron scattering on an atom is given. This proof does not
require any assumptions about the dependence of the scattering phases on
energy. For electron energies lower than the
luminescence threshold, it is shown that the predictions for a number of radiated photons obtained by the exact formula are in good agreement
with the available experimental data.
\end{abstract}
\maketitle

\section{Introduction}

In recent years, search for particles of dark matter has stimulated the development of
detectors with increased sensitivity to registration of
such particles~\citep{Buzulutskov2018,Bondar2020a,Borisova2021,Borisova2022,Henriques2022}.
One of the fundamental processes, that determine the properties of new
detectors, is  bremsstrahlung on noble gases, in particular
on argon or xenon. Many years ago approximate
formulas for describing the bremsstrahlung process at low energies were suggested~\citep{Ohmura1960,Firsov1960}.
In these formulas, the bremsstrahlung spectrum is expressed in terms of the electron scattering cross section on an atom. These approximate formulas were used to
 compare experimental data with theory. In Ref.~\citep{Biberman1967}
it was noted that the region of applicability of the approximate formulas is not defined.
In particular, it is not clear whether these formulas can be used to describe
bremsstrahlung spectrum on noble gases, since in this case the Ramsauer effect~\citep{Ramsauer1922} appears.

In our work, we carry out a detailed comparison of the bremsstrahlung spectra
of nonrelativistic electron on argon and xenon obtained by the exact and approximate
formulas. Exact formulas have been known for a long time (see, e.g., Refs.~\citep{Ashkin1966,Dyachkov1983} and references therein). In the limit
when the photon frequency $\omega$ tends to zero, the bremsstrahlung spectrum
 is expressed in terms of the transport cross section in the process of elastic
scattering of an electron on an atom. However, the proof of this statement for
 nonrelativistic electrons was obtained only in various approximations~\citep{Kasyanov1965,landau4,Amusia1988,Amusia1990}.
For instance, the Born approximation or the approximation in which the
contribution of the $s$-wave to the scattering cross section is dominant were discussed.
We present  for the first time a
rigorous proof of the relationship between the bremsstrahlung spectrum and the transport
cross section at $\omega\to0$ and arbitrary energies of the incident electron
(compared to the energies of atomic electrons). In addition, it is shown that the approximate formulas used in the literature for the bremsstrahlung spectrum at finite photon frequencies give
predictions that differ significantly from the exact results.
We use the results for the spectrum for comparison of the predictions for a number of emitted photons in a certain wavelength region with the experimental results~\citep{Buzulutskov2018,Bondar2020a,Borisova2021,Borisova2022,Henriques2022}.

\section{Theory}
Let us consider the process of bremsstrahlung at the scattering of a nonrelativistic
electron with kinetic energy $\varepsilon$ on an atom. For an electron wave function, we use the partial wave expansions
\begin{equation}
\psi^{(\pm)}(\boldsymbol{r})=\frac{1}{2p}\sum_{l=0}^{\infty}\left(2l+1\right)i^l e^{\pm i\delta_{l}}R_{l}(p,r)P_{l}(\cos\theta)\,,\label{eq:psi}
\end{equation}
where $p=\sqrt{2m\varepsilon}$ is the electron momentum, $\delta_{l}$ are the scattering phases, $P_{l}$ are the Legendre polynomials, and $R_{l}(p,r)$ are the
radial wave functions having asymptotics at large distances
\begin{equation}
R_{l}(p,r)\xrightarrow{r\to\infty}\frac{2}{r}\sin\left(pr-\frac{\pi l}{2}+\delta_{l}\right).\label{eq:Ras}
\end{equation}
The function $\psi^{(+)}(\bm r)$ contains at large distances the plane wave and a divergent spherical wave, while $\psi^{(-)}(\bm r)$ contains the plane wave and a convergent spherical wave.
Then, the bremsstrahlung spectrum in the nonrelativistic approximation has the form (see, e.g., Ref.~\citep{Dyachkov1983} and Appendix)
\begin{align}
 & \frac{d\sigma}{d\omega}=\frac{2\alpha}{3\omega}\frac{p_{f}}{p_{i}}\sum_{l=0}^{\infty}\left(l+1\right)\left(|M_{l,l+1}(p_{f},p_{i})|^2+|M_{l+1,l}(p_{f},p_{i})|^2\right),\nonumber \\
 & M_{l',l}(p_{f},p_{i})=\frac{\exp{\{i[\delta_l(p_i)+\delta_{l'}(p_f)]\}}}{p_{i}p_{f}}\int_0^\infty r^{2}dr\,R_{l'}(p_{f},r)\frac{\partial U}{\partial r}R_{l}(p_{i},r)\,,\label{eq:brsection}
\end{align}
where $\alpha$ is the fine-structure constant, $\omega=\varepsilon_{i}-\varepsilon_{f}$ is the emitted photon frequency, $\varepsilon_{i}$ and $\varepsilon_{f}$ are the
electron energies before and after collision, $U(r)$ is the electron potential energy in an atomic field, $\hbar=c=1$. Below we will refer Eq.~\eqref{eq:brsection} as the exact formula in contrast to various approximations to this result.

For $\omega\ll\varepsilon_{i}$ the expression~\eqref{eq:brsection} is
noticeably simplified:
\begin{equation}
\frac{d\sigma}{d\omega}=\frac{4\alpha}{3\omega}\sum_{l=0}^{\infty}\left(l+1\right)|M_{l,l+1}(p_{i},p_{i})|^2\,.
\end{equation}
The matrix element
\begin{equation}\label{mat0}
T_{l}(p)=M_{l,l+1}(p,p)=\frac{\exp{\{i[\delta_l(p)+\delta_{l+1}(p)]\}}}{p^{2}}\int_0^\infty r^{2}dr\,R_{l}(p,r)\frac{\partial U}{\partial r}R_{l+1}(p,r)
\end{equation}
can be expressed in terms of the scattering phases $\delta_{l}$ for any momenta
$p$. To prove this statement, we use the relations
\begin{align}
 & \frac{\partial U}{\partial r}=i\left[\mathcal{P}_{r},H_{l}\right]+\frac{l(l+1)}{mr^{3}}\,,\nonumber \\
 & H_{l+1}-H_{l}=\frac{l+1}{mr^{2}}\,,\nonumber \\
 & H_{l}=\frac{\mathcal{P}_{r}^{2}}{2m}+U(r)+\frac{l(l+1)}{2mr^{2}}\,,
\end{align}
where $\mathcal{P}_{r}=-i\left(\frac{1}{r}+\frac{\partial}{\partial r}\right)$ is
the radial momentum operator, $H_{l}$ is the radial Hamiltonian for an
electron in a state with orbital momentum~$l$, and $m$ is the electron mass. However, it is impossible to use the hermiticity of the Hamiltonian in Eq.~\eqref{mat0} and make the replacement
\begin{equation}
    \int_0^\infty r^{2}dr\,R_{l}(p,r)\left[\mathcal{P}_{r},H_{l}\right]R_{l+1}(p,r) \Longrightarrow
    -\int_0^\infty r^{2}dr\,R_{l}(p,r)\,\mathcal{P}_{r}\,\frac{l+1}{mr^{2}}\,R_{l+1}(p,r)\,,\nonumber
\end{equation}
 since the integral is conditionally convergent. To overcome this difficulty, we use the trick described in~\citep{Sushkov2013,milstein2013kinetics}.
Consider the regularized matrix element
\begin{equation}
\widetilde{T}_{l}(p)=\frac{\exp{\{i[\delta_l(p)+\delta_{l+1}(p)]\}}}{p^{2}}\int_0^\infty r^{2}dr\,R_{l}(p,r)\frac{\partial U}{\partial r}R_{l+1}(p,r)\,e^{-\lambda r}\,,
\end{equation}
where $\lambda\to0$. In the expression for $\widetilde{T}_{l}(p)$,
 the hermiticity of the Hamiltonian can already be used. Neglecting the terms quadratic in $\lambda$ in pre-exponent and still keeping $e^{-\lambda r}$ in the integrand, we reduce the matrix element to the form
\begin{align}
 & \widetilde{T}_{l}(p)=\exp{\{i[\delta_l(p)+\delta_{l+1}(p)]\}}\cdot\left(t_{1}+t_{2}\right),\nonumber \\
 & t_{1}=\frac{l+1}{2mp^{2}}\int_0^\infty dr\,e^{-\lambda r}\left[ R_{l}'R_{l+1}-R_{l}R_{l+1}'+\frac{2(l+1)}{r}\,R_{l}R_{l+1}\right] ,\nonumber \\
 & t_{2}=-\frac{\lambda}{p^{2}}\int_0^\infty r^{2}dr\,e^{-\lambda r}\left[ R_{l}\frac{\mathcal{P}_{r}^{2}}{2m}R_{l+1}-R_{l+1}\frac{\mathcal{P}_{r}^{2}}{2m}R_{l}\right] ,\label{eq:Mel}
\end{align}
where $R_L'=\partial R_L/\partial r$. Note that the factor $e^{-\lambda r}$ in the integrand
for $t_{1}$ can be omitted
since the integral is already absolutely convergent. Using the radial
Schr\"odinger equation for the wave functions $R_{l}$ and $R_{l+1}$, we obtain
the relation
\begin{equation}
R_{l}'R_{l+1}-R_{l+1}'R_{l}+\frac{2(l+1)}{r}\,R_{l}R_{l+1}=\frac{\partial}{\partial r}\left[r\left(R_{l+1}'R_{l}-R_{l}'R_{l+1}\right)\right].
\end{equation}
Therefore, the contribution of $t_{1}$ vanishes,
\begin{equation}
t_{1}=\frac{l+1}{2mp^{2}}\int_0^\infty dr\left[R_{l}'R_{l+1}-R_{l}R_{l+1}'+\frac{2(l+1)}{r}\,R_{l}R_{l+1}\right]=0\,.\label{eq:t1}
\end{equation}
Since the matrix element $t_{2}$ is proportional to the small factor
$\lambda$, this factor can only be compensated by the
contribution to the integral of large distances $r$. Therefore, we can use the asymptotics~\eqref{eq:Ras} of the wave functions and replace $\frac{\mathcal{P}_{r}^{2}}{2m}$
by~$\varepsilon$. As a result, in the limit $\lambda\to0$ we obtain
\begin{align}
& t_{2}=-\frac{8\lambda\,\varepsilon}{p^{2}}\int_0^\infty dr\,e^{-\lambda r}\sin\left(pr-\frac{\pi l}{2}+\delta_{l}\right)\sin\left(pr-\frac{\pi(l+1)}{2}+\delta_{l+1}\right) \nonumber \\
& =\frac{2}{m}\sin\left(\delta_{l}-\delta_{l+1}\right).\label{eq:t2}
\end{align}
Hence, the matrix element~\eqref{mat0} reads
\begin{equation}
	T_l(p)=\frac{2}{m}\sin\left(\delta_{l}-\delta_{l+1}\right)\exp{\{i[\delta_l(p)+\delta_{l+1}(p)]\}}.\label{eq:Tl}
\end{equation}

Using Eq.~\eqref{eq:Tl}, we arrive at the expression
for the asymptotics of the bremsstrahlung spectrum at $\omega\to0$:
\begin{equation}
\frac{d\sigma}{d\omega}=\frac{16\alpha}{3\omega m^{2}}\sum_{l=0}^{\infty}\left(l+1\right)\sin^{2}\left(\delta_{l}-\delta_{l+1}\right).
\end{equation}
This formula can be expressed in terms of the transport scattering cross section $\sigma_{\mathrm{tr}}$
 of an electron on an atom,
\begin{equation}
\sigma_{\mathrm{tr}}=\frac{4\pi}{p^{2}}\sum_{l=0}^{\infty}\left(l+1\right)\sin^{2}\left(\delta_{l}-\delta_{l+1}\right).
\end{equation}
Finally, we obtain the bremsstrahlung spectrum for low frequencies
\begin{equation}
\frac{d\sigma}{d\omega}=\frac{8\alpha}{3\pi\omega}\frac{\varepsilon}{m}~\sigma_{\mathrm{tr}}\,.\label{eq:brsectionasfinal}
\end{equation}

\begin{figure}[!tb]
\includegraphics[totalheight=5.6cm]{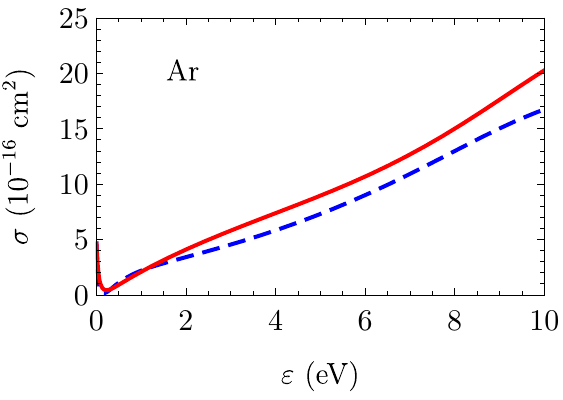}\hfill{}\includegraphics[totalheight=5.6cm]{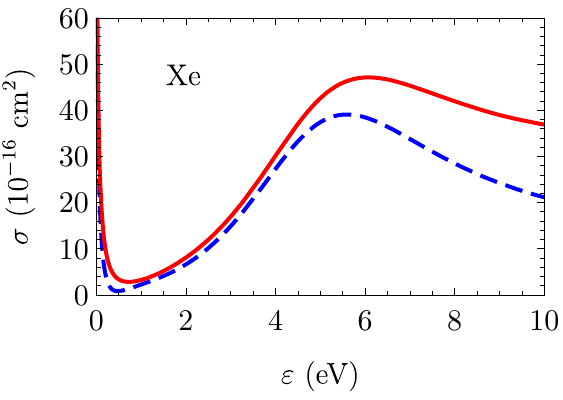}

\caption{Cross sections for electron scattering on argon and xenon atoms calculated
	using the potential~\eqref{eq:pot}. The solid line corresponds to the
	elastic scattering cross section $\sigma_{\mathrm{el}}$, and the dashed line corresponds to the 	transport cross section $\sigma_{\mathrm{tr}}$.}\label{fig:crosssections}
\end{figure}

In some papers (see~\citep{Biberman1967,Amusia1990} and references therein), to describe the bremsstrahlung spectrum for arbitrary
frequencies $\omega$,  the formula
\begin{equation}
\frac{d\sigma}{d\omega}=\frac{4\alpha}{3\pi m\omega}\frac{p_{i}}{p_{f}}\left[\varepsilon_{i}\sigma(\varepsilon_{f})+\varepsilon_{f}\sigma(\varepsilon_{i})\right],\label{eq:brsectioncommon}
\end{equation}
was applied. This formula coincides with~\eqref{eq:brsectionasfinal} in the limit $\omega\to0$,
if the transport cross section is used as $\sigma$. Similar
expression in terms of the scattering cross section is widely used to describe the
cross section of photon absorption which is the inverse process to bremsstrahlung.
Besides, in most of cases the authors used only the contribution of  $s$-wave \citep{Ohmura1960, Somerville1964, Dalgarno1966, Biberman1967},
or used as $\sigma$ the cross section of elastic scattering by an
atom~\citep{Firsov1960,Buzulutskov2018,Bondar2020a,Borisova2021,Borisova2022},
\begin{equation}
\sigma_{\mathrm{el}}=\frac{4\pi}{p^{2}}\sum_{l=0}^{\infty}\left(2l+1\right)\sin^{2}\delta_{l}\,.
\end{equation}
In Refs.~\citep{Kasyanov1965,Henriques2022} the relationship \eqref{eq:brsectioncommon} between the bremsstrahlung spectrum and the transport cross section was pointed out, but
the authors of these works supposed that using of an elastic cross section is also
admissible. However, the cross sections $\sigma_{\mathrm{el}}$ and $\sigma_{\mathrm{tr}}$
in the case of noble gases differ significantly even for the energy of the incident
electron  much less than the energy of atomic electrons (see Fig.~\ref{fig:crosssections}).
Note that the minimum in the cross sections is a consequence of the Ramsauer effect, which
manifests in a nontrivial energy dependence of the scattering phase $\delta_{0}$ at low-energy electron scattering on noble gases~\citep{Bell1984,Plenkiewicz1988,Saha1991,Endredi1994,McEachran1997,Johnson1994,Gibson1998}.
For example, $\delta_{0}$ vanishes at the electron energy $\unit[0.3]{eV}$
in the case of argon and $\unit[0.8]{eV}$ in the case of xenon.

\section{Results and discussion}

For a quantitative description of the bremsstrahlung spectrum and the electron scattering cross section in the cases of argon and xenon, we use the potential energy $U(r)$ as a sum
\begin{align}
 & U(r)=U_{\mathrm{st}}(r)+U_{\mathrm{pol}}(r)\,,\nonumber \\
 & U_{\mathrm{pol}}(r)=-\left[\frac{\alpha_{d}}{\left(\rho^{2}+d^{2}\right)^{2}}+\frac{\alpha_{q}}{\left(\rho^{3}+d^{3}\right)^{2}}\right]\mathrm{Ry}\,,\qquad\rho=\frac{r}{a_{B}}\,,\label{eq:pot}
\end{align}
where $U_{\mathrm{st}}$ is a static potential determined by the charge distribution
 in the atom, $U_{\mathrm{pol}}$ is a polarization potential,
$\mathrm{Ry}=me^{4}/2\hbar^{2}$, $a_{B}=\hbar^{2}/me^{2}$, $e$ is the electron
charge. The values of the dimensionless parameters of the polarization potential
$\alpha_{d}$, $\alpha_{q}$ and $d$ used in our paper provide
agreement between theoretical predictions for the elastic electron scattering cross section with experimental data~\citep{Kurokawa2011}. These values are given in the Table \ref{tab:PolPot}. Note that an account for the polarization potential is very important for correct description of the scattering cross section in the case of noble gases due to large values of the parameters $\alpha_{d}$ and~$\alpha_{q}$.
The static potential $U_{\mathrm{st}}$ was calculated using the Hartree-Fock-Dirac method~\citep{Bratsev1977}.

\setlength{\tabcolsep}{1.5em}

\begin{table}[!tb]
\begin{centering}
\begin{tabular}{|l|c|c|c|}
\hline
 & $\alpha_{d}$ & $\alpha_{q}$ & $d$\tabularnewline
\hline
Ar & $13.9$ & $60.5$ & $1.95$\tabularnewline
\hline
Xe & $30.5$ & $134.7$ & $2.13$\tabularnewline
\hline
\end{tabular}
\par\end{centering}
\caption{The dimensionless parameters of the polarization potential~\eqref{eq:pot}.}\label{tab:PolPot}
\end{table}

\begin{figure}[!tb]
\includegraphics[totalheight=5.6cm]{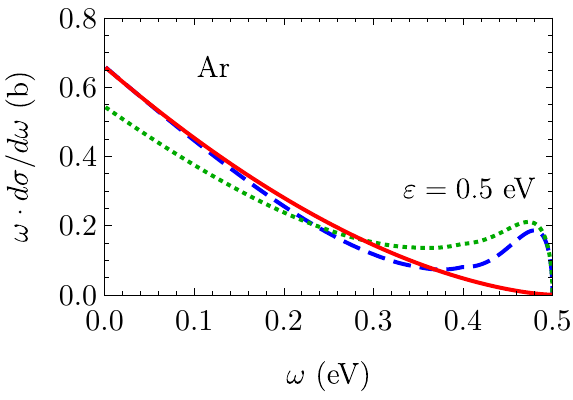}\hfill{}\includegraphics[totalheight=5.6cm]{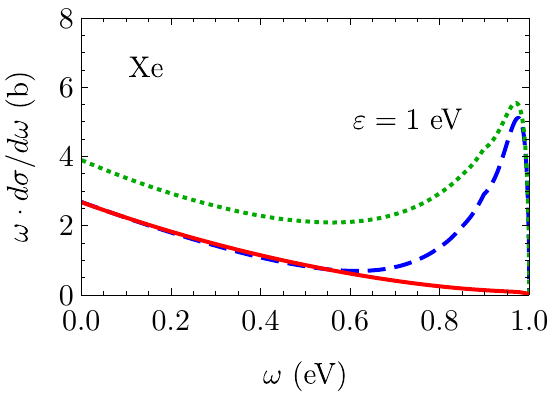}

\includegraphics[totalheight=5.6cm]{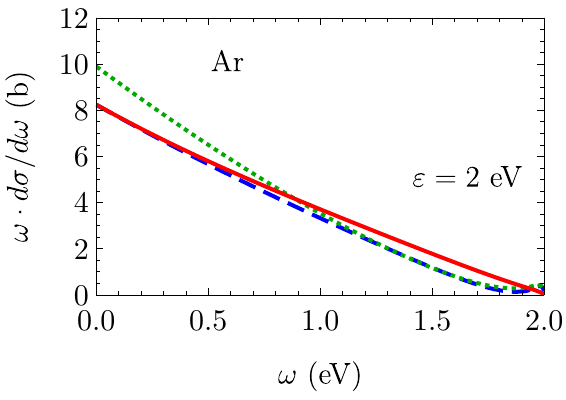}\hfill{}\includegraphics[totalheight=5.6cm]{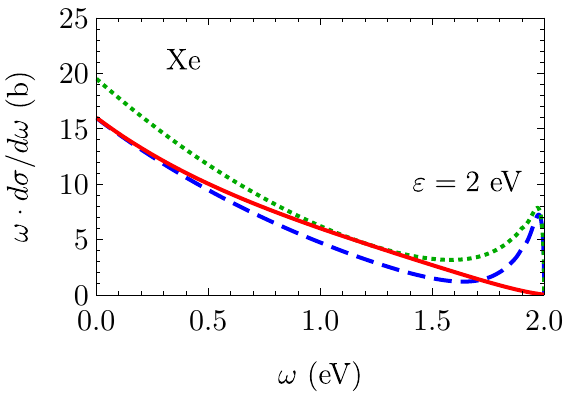}

\includegraphics[totalheight=5.6cm]{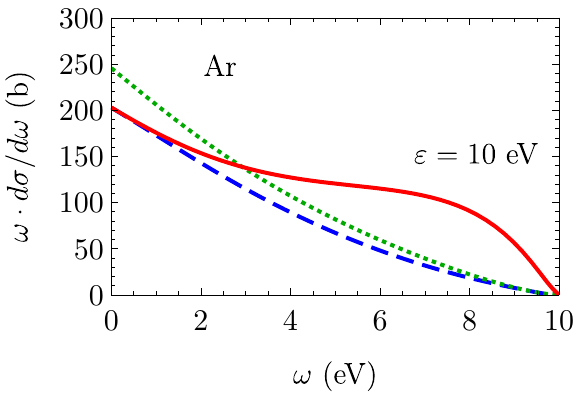}\hfill{}\includegraphics[totalheight=5.6cm]{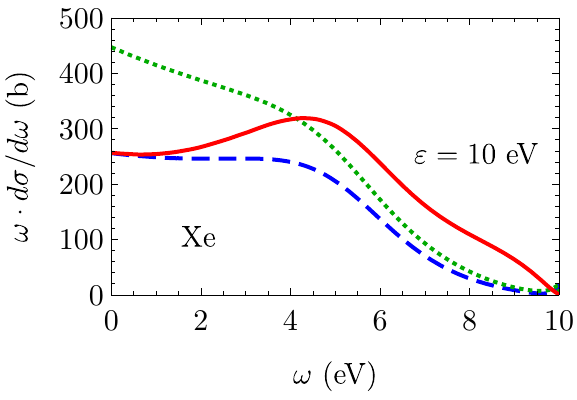}

\caption{The frequency dependence of $\omega\cdot d\sigma/d\omega$
	for scattering of electrons on argon and xenon atoms.
	The solid line corresponds to the exact formula~\eqref{eq:brsection},
	the dashed line corresponds to Eq.~\eqref{eq:brsectioncommon}
	expressed in terms of the transport cross section $\sigma_{\mathrm{tr}}$, while the dotted
	line corresponds to Eq.~\eqref{eq:brsectioncommon} expressed through~$\sigma_{\mathrm{el}}$.}\label{fig:spectres}
\end{figure}

The bremsstrahlung spectra
on argon and xenon, obtained by means of the exact~\eqref{eq:brsection}
and approximate~\eqref{eq:brsectioncommon} formulas, are shown in Fig.~\ref{fig:spectres} for energies
$\varepsilon=\unit[0.5\div10]{eV}$. It is seen that the applicability of the approximate
formulas for the description of the bremsstrahlung spectrum in the specified energy region
 is very limited. Note that at low photon frequencies
agreement of the exact formula with the approximate formula based on the transport cross section $\sigma_{\mathrm{tr}}$ is much better than with the approximate
formula based on~$\sigma_{\mathrm{el}}$.

\begin{figure}[!b]
\includegraphics[totalheight=5.6cm]{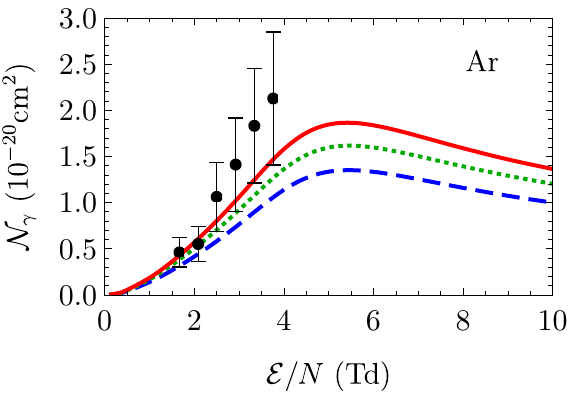}\hfill{}\includegraphics[totalheight=5.6cm]{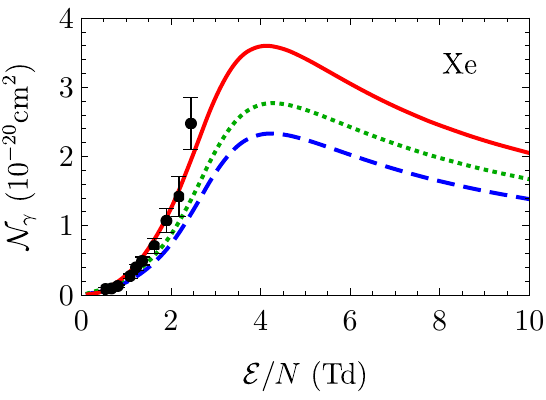}

\caption{The reduced yield of bremsstrahlung photons as a function of the reduced
	electric field for argon and xenon. Theoretical predictions are
	obtained by means of Eq.~\eqref{eq:Yield}. The solid line corresponds to the exact formula~\eqref{eq:brsection} for $d\sigma/d\omega$,
	the dashed line corresponds to Eq.~\eqref{eq:brsectioncommon}
	expressed in terms of the transport cross section $\sigma_{\mathrm{tr}}$, while the dotted
	line corresponds to Eq.~\eqref{eq:brsectioncommon} expressed
	via~$\sigma_{\mathrm{el}}$. Experimental data for argon are
	taken from Ref.~\citep{Borisova2021}, and data for xenon are recalculated
	from that in Ref.~\citep{Henriques2022}.}\label{fig:Yields}
\end{figure}

In Refs.~\citep{Buzulutskov2018,Bondar2020a,Borisova2021,Borisova2022,Henriques2022}
experimental data for the number of emitted photons are given
for electrons accelerated in an electric field and scattered on
argon or xenon atoms. In this case, photons were registered in the
wavelength region $\lambda=\unit[0\div1000]{nm}$ for argon and $\lambda=\unit[120\div1000]{nm}$
for xenon. Reduced yield of bremsstrahlung photons $\mathcal{N}_{\gamma}$,
which is defined as the number of bremsstrahlung photons per electron
per atomic concentration and per drift path, is given by
\begin{equation}
\mathcal{N}_{\gamma}=\int_{\lambda_{\mathrm{min}}}^{\lambda_{\mathrm{max}}}d\lambda\int_{\omega}^{\infty}d\varepsilon\,\frac{v_{e}}{v_{d}}\frac{d\sigma}{d\omega}\frac{d\omega}{d\lambda}\,f(\varepsilon)\,,\label{eq:Yield}
\end{equation}
where $v_{e}=\sqrt{2\varepsilon/m}$ is the electron velocity, $v_{d}$ is the drift velocity, $f(\varepsilon)$ is the electron distribution function normalized as
\begin{equation}
\int_{0}^{\infty}d\varepsilon\,f(\varepsilon)=1\,.
\end{equation}
The electron distribution function and the drift velocity are determined by
the magnitude of the electric field. For these values, we used the results
obtained by means of EEDF~\citep{Dyatko2011}. The dependence of $\mathcal{N}_{\gamma}$
on the ratio $\mathcal{E}/N$, where $\mathcal{E}$ is the electric
field and $N$ is the concentration of atoms, is shown in Fig.~\ref{fig:Yields}.
For the reduced electric field, we use conventional units
$\unit[1]{Td}=\unit[10^{-17}]{V\cdot cm^{2}}$.
It is seen that the predictions
for $\mathcal{N}_{\gamma}$, obtained from the exact formulas, significantly
differ from those obtained using approximate formulas. Note that
this difference essentially depends on the region of integration over wavelengths  in Eq.~\eqref{eq:Yield}. The experimental data for the photon yield at values
$\mathcal{E}/N$, which lead to emission of photons below the threshold of electroluminescence, are also shown In Fig.~\ref{fig:Yields}.
 Above this threshold, the photon yield due to
electroluminescence significantly exceeds the photon yield due to
bremsstrahlung. It is important that the predictions obtained from the exact
formula are in much better agreement with the experimental data than
obtained by approximate formulas. This is especially noticeable in the case of
xenon.

\section{Influence of polarization radiation}


When nonrelativistic electrons are scattered by an atom without excitation of atomic electrons, the emission of photons is related not only to the radiation of the incident electron (bremsstrahlung) but also to the radiation of atomic electrons in the intermediate states (polarization radiation, see \citep{Amusia1988,Amusia1990}). In this case, if the energy of the incident electron is comparable to the interval between atomic levels, then the phenomenon of electroluminescence occurs. The intensity of radiation due to electroluminescence significantly exceeds the intensity of bremsstrahlung. Therefore, it is important to discuss the effect of polarization radiation on the photon spectrum below the electroluminescence threshold.

\begin{figure}[!tb]
    \includegraphics[totalheight=5.6cm]{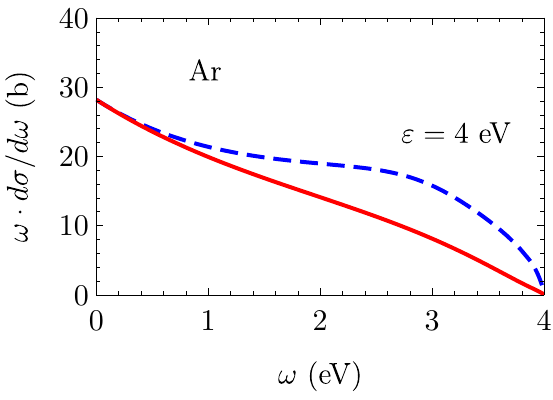}\hfill{}\includegraphics[totalheight=5.6cm]{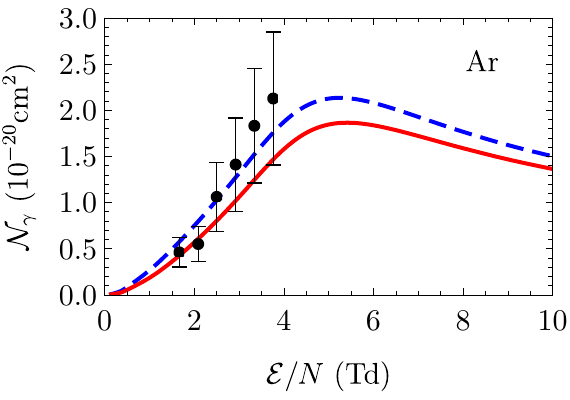}

    \caption{The frequency dependence of $\omega\cdot d\sigma/d\omega$
        for scattering of electrons with $\varepsilon=\unit[4]{eV}$ on argon (left figure) and the reduced photon yield as a function of the reduced
        electric field (right figure).
        The solid lines correspond to the matrix element given in Eq.~\eqref{eq:brsection} and
        the dashed line to that given in Eq.~\eqref{eq:polrad}.
        Experimental data are
        taken from Ref.~\citep{Borisova2021}.\label{fig:Polrad}}
\end{figure}

Assumimg that the main contribution to the amplitude of polarization rediation is given by the motion of incident electron at large distances compared to an atomic size, the total amplitude of the photon emission below the threshold of electroluminescence can be written as
\begin{equation}
M_{l',l}(p_{f},p_{i})=\frac{\exp{\{i[\delta_l(p_i)+\delta_{l'}(p_f)]\}}}{p_{i}p_{f}}\int_0^\infty r^{2}dr\,R_{l'}(p_{f},r)\left[\frac{\partial U}{\partial r}-\frac{\alpha_{d}\,a_B^3 m\omega^2}{r^2}\right]R_{l}(p_{i},r)\,\label{eq:polrad},
\end{equation}
where $\alpha_{d}$ is the static polarizability which coincides with that given in the Table~\ref{tab:PolPot}. In Eq.~\eqref{eq:polrad} $\omega$ is small compared to the resonant frequency (luminescence frequency). As should be, at $\omega\to0$ the contribution of polarization radiation vanishes. Since the contribution of polarization radiation in Eq.~\eqref{eq:polrad} was obtained under the assumption $r\gg a_B$, account for the distances of the order $r\sim a_B$ requires a special consideration. For a qualitative discussion, we replace $a_B^2/r^2$ in the polarization contribution by $1/(\rho^2+d^2)$ in the same way as in Eq.~\eqref{eq:pot}, where $\rho=r/a_B$. Comparison of theoretical predictions with and without account for the polarization radiation and the experimental data is given in Fig.~\ref{fig:Polrad} for the case of argon. It is seen that account for the polarization radiation in the near-threshold region noticeably improves the agreement between theory and experiment.


\section{Photon angular distribution}

From the experimental point of view, it is also interesting to consider the angular distribution of the radiated photons. For this quantity, the exact expression reads (see Appendix)
\begin{align}
	&d\sigma_{\gamma}=\frac{d\sigma}{d\omega}\cdot\frac{d\omega\,d\Omega_{\bm k}}{4\pi}\cdot\left[1+\beta\,P_2(\cos \theta_{\bm{k}})\right],\qquad P_2(y)=\dfrac{3y^2-1}{2}\,,\nonumber \\
	&\beta=\frac{1}{2}\left(\sum_{l=0}^\infty (l+1) \left[|M_{l,l+1}|^2+|M_{l+1,l}|^2\right]\right)^{-1}\nonumber \\
	&\times\sum_{l=0}^\infty (l+1)\left\{\dfrac{6(l+2)}{2l+3}\,
	\mbox{Re}[M_{l+1,l}M^{*}_{l+1,l+2}]-\dfrac{l+2}{2l+1}|M_{l,l+1}|^2-\dfrac{l}{2l+3}|M_{l+1,l}|^2\right\},
\end{align}
where $\theta_{\bm{k}}$ is the angle between the momentum of the initial electron and the photon momentum, $\frac{d\sigma}{d\omega}$ is given in~\eqref{eq:brsection}, and $M_{l',l}$ are given in~\eqref{eq:polrad}. Using Eqs.~\eqref{mat0} and~\eqref{eq:Tl} we find the asymmetry $\beta$ at $\omega \ll \varepsilon_i$
\begin{align}
	& \beta(\omega \ll \varepsilon_i)=\frac{1}{4} \left(\sum_{l=0}^\infty (l+1) \sin^2(\delta_l-\delta_{l+1})\right)^{-1} \nonumber \\
	& \times \sum_{l=0}^\infty (l+1)\sin(\delta_l-\delta_{l+1})\left\{\frac{3(l+2)}{2l+3}\sin(\delta_l+\delta_{l+1}-2\delta_{l+2})-\frac{5l+4}{2l+1}\sin(\delta_l-\delta_{l+1})\right\}.
\end{align}

\begin{figure}[!tb]
	\includegraphics[totalheight=5.5cm]{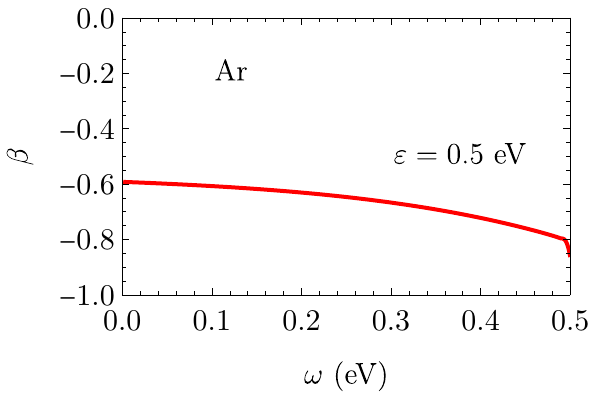}\hfill{}\includegraphics[totalheight=5.5cm]{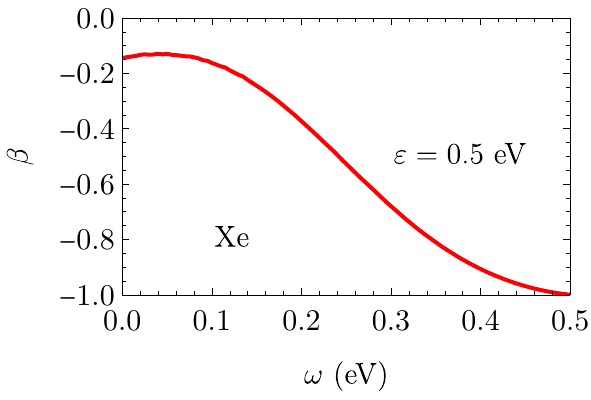}
	
	\includegraphics[totalheight=5.5cm]{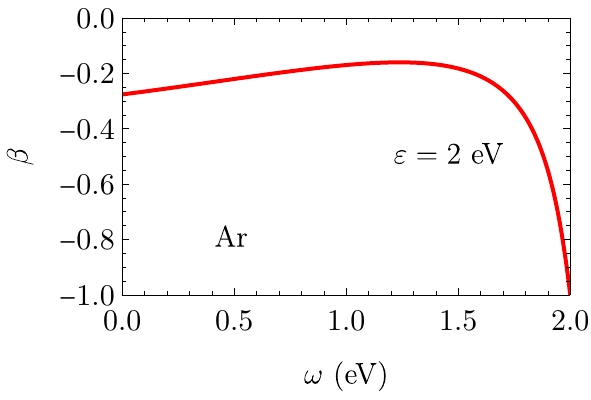}\hfill{}\includegraphics[totalheight=5.5cm]{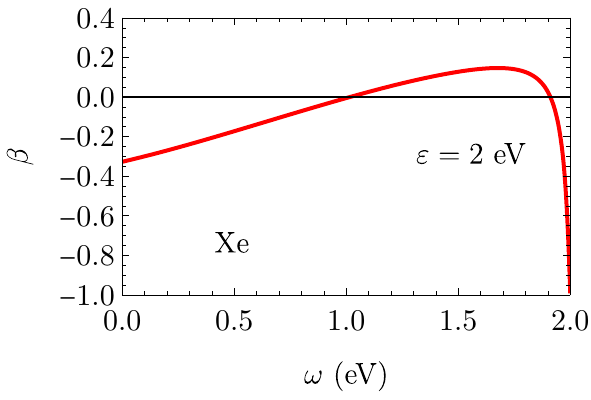}
	
	\includegraphics[totalheight=5.5cm]{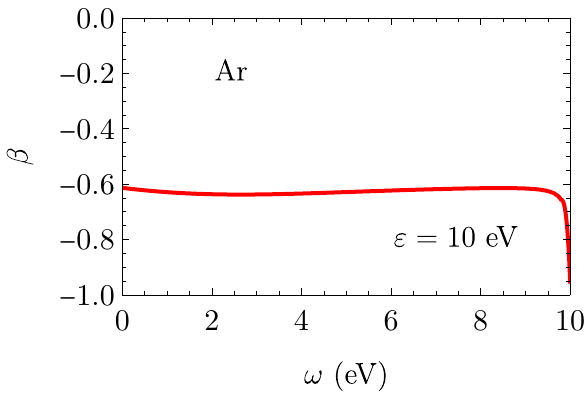}\hfill{}\includegraphics[totalheight=5.5cm]{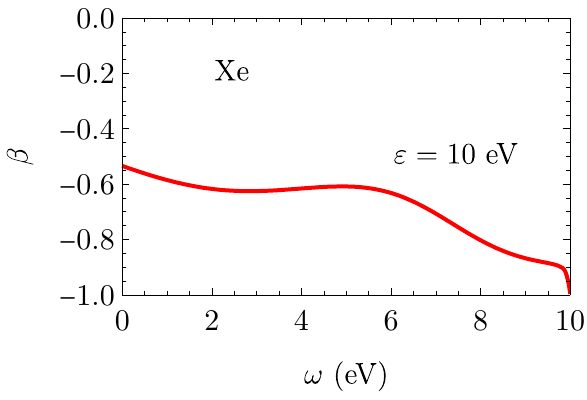}
	
	\caption{The photon energy dependence of the asymmetry $\beta$ for bremsstrahlung on argon and xenon at a few electron energies.}\label{fig:angular}
\end{figure}

Our predictions for the asymmetry $\beta$ for bremsstrahlung on argon and xenon are shown in Fig.~\ref{fig:angular}. It is seen that the asymmetry is negative for almost all photon and electron energies considered. In the limit $\omega\to\varepsilon_i$ the asymmetry tends to $\beta\to-1$, which can be explained as follows. The matrix element $\mathcal{M}$ of the process can be written as $\mathcal{M}=\bm{e}\cdot\bm{J}$, where $\bm{e}$ is the photon polarization vector and the vector $\bm{J}$ is expressed via the momenta $\bm{p}_i$ and $\bm{p}_f$ of initial and final electrons (see Appendix). Summation over the photon polarizations gives $$\sum |\mathcal{M}|^2=|\bm{J}|^2-|\bm{n}_{\bm k}\cdot\bm{J}|^2\,,$$ where $\bm{n}_{\bm k}=\bm{k}/k$ and $\bm{k}$ is the photon momentum. If $\omega=\varepsilon_i$ then $\bm{p}_f=0$ and $\bm{J}\propto \bm{p}_i$. Therefore, $\sum|\mathcal{M}|^2 \propto 1-\cos^2 \theta_{\bm{k}}$, which corresponds to $\beta=-1$.

\section{Conclusion}

In this work, we have carried out a detailed analysis of the bremsstrahlung spectrum at scattering of nonrelativistic electrons on argon and xenon. Predictions obtained by the exact formula~\eqref{eq:brsection}
differ significantly from the predictions obtained by means of the approximate
formula~\eqref{eq:brsectioncommon} widely used in the literature.
This statement is true both for approximate formulas expressed via the
transport cross section $\sigma_{\mathrm{tr}}$ and for formulas expressed via~$\sigma_{\mathrm{el}}$. It is shown that the predictions for the photon  yield  below the luminescence threshold, obtained by means of the exact formula, are in good
agreement with the available experimental data.

In the limit $\omega\to0$, a rigorous proof is given for the
relation between the bremsstrahlung spectrum and the transport scattering cross section.
Unlike previous works, this proof does not require any
assumptions on the energy dependence of scattering phases.

\section*{Acknowledgements}
	We are grateful to A.F. Buzulutskov and E.A. Frolov for valuable discussions.


\section*{Appendix}

In this Appendix we derive the exact formulas for the angular distributions in the process of bremsstrahlung. The differential cross section has the form
\begin{align}
	& d\sigma = \frac{\alpha}{\omega} \frac{p_f}{p_i}\frac{d\omega\,d\Omega_{{\bm p}_f}d\Omega_{\bm k}}{(2\pi)^4}\,|\mathcal{M}|^2,\qquad \mathcal{M}=\bm e\cdot\bm J \,, \nonumber \\
	& \bm{J}=i \int \psi_f^{(-)*}(\bm{r})\, \bm{n}\, \frac{\partial U}{\partial r}\, \psi_i^{(+)}(\bm{r})\,,\qquad \bm{n}=\frac{\bm r}{r}\,,
	\label{eq:angular}
\end{align}	
where the wave functions $\psi^{(\pm)}(\bm r)$ are given in~\eqref{eq:psi}. The matrix element $\bm{J}$ is expressed via the vectors $\bm{\lambda}_i=\bm{p}_i/p_i$ and $\bm{\lambda}_f=\bm{p}_f/p_f$, so that we can write $\bm{J}$~as
 $$\bm J=\frac{\bm\lambda_i+\bm\lambda_f}{1+x}\,A + \frac{\bm\lambda_i-\bm\lambda_f}{1-x}\,B\,,\qquad x=\bm\lambda_i\cdot\bm\lambda_f\,,$$
 where $$A=\frac{1}{2}(\bm\lambda_i+\bm\lambda_f)\cdot\bm{J}\,,\qquad B=\frac{1}{2}(\bm\lambda_i-\bm\lambda_f)\cdot\bm{J}\,.$$
 Using the recurrent relation for the Legendre polynomials $$x P_l(x)=\frac{(l+1)P_{l+1}(x)+l P_{l-1}(x)}{2l+1}$$ and the orthogonality relation of the Legendre polynomials we obtain
\begin{align}
	& A=\frac{\pi}{2}\sum_{l=0}^\infty (l+1) \left[P_l(x)+P_{l+1}(x)\right]\left[M_{l,l+1}-M_{l+1,l}\right],\nonumber \\
	& B=\frac{\pi}{2}\sum_{l=0}^\infty (l+1) \left[P_l(x)-P_{l+1}(x)\right]\left[M_{l,l+1}+M_{l+1,l}\right],
\end{align}
where $M_{l',l}$ are given in~\eqref{eq:polrad}.

After summation over the photon polarizations and integration over $\Omega_{\bm{k}}$ we obtain the electron angular distribution
\begin{align}
	&d\sigma_{e}=\frac{\alpha}{3\pi^3\omega} \dfrac{p_i}{p_f}\,d\omega\,d\Omega_{{\bm p}_f}\left[\dfrac{|A|^2}{1+x}+\dfrac{|B|^2}{1-x}\right].
\end{align}

To obtain the photon angular distribution we write 
\begin{align}
	& \int d\Omega_{{\bm p}_f}\,J^a J^{*b}=a\,\delta^{ab}+b\,(\delta^{ab}-3\lambda^a_i\lambda^b_i) \,, \nonumber\\
	& a = \frac{1}{3}\int d\Omega_{{\bm p}_f}\,|\bm J|^2\,,\qquad b=\frac{a}{2}-\frac{1}{2}\int d\Omega_{{\bm p}_f}\,|\bm{\lambda}_i\cdot\bm J|^2\,.
\end{align}
Then we use the relations
\begin{align}
	& \int_{-1}^1 dx \,\frac{[P_{l+1}(x)+P_l(x)][P_{l'+1}(x)+P_{l'}(x)]}{1+x}=\frac{2\delta_{ll'}}{l+1}\,, \nonumber \\
	& \int_{-1}^1 dx \,\frac{[P_{l+1}(x)-P_l(x)][P_{l'+1}(x)-P_{l'}(x)]}{1-x}=\frac{2\delta_{ll'}}{l+1}\,,
\end{align}
which can easily be proved. Finally, we have the photon angular distribution
\begin{align}
	&d\sigma_{\gamma}=\frac{d\sigma}{d\omega}\cdot\frac{d\omega\,d\Omega_{\bm k}}{4\pi}\left[1+\beta\,P_2(\bm\lambda_i\cdot{\bm n}_{\bm k})\right],\qquad P_2(y)=\dfrac{3y^2-1}{2}\,,\qquad \beta=\frac{b}{a}\,,\nonumber \\
	&a=\frac{4\pi^3}{3}\sum_{l=0}^\infty (l+1) \left[|M_{l,l+1}|^2+|M_{l+1,l}|^2\right], \nonumber \\
	&b=\frac{2\pi^3}{3}\sum_{l=0}^\infty (l+1)\left\{\dfrac{6(l+2)}{2l+3}\,
	\mbox{Re}[M_{l+1,l}M^{*}_{l+1,l+2}]-\dfrac{l+2}{2l+1}|M_{l,l+1}|^2-\dfrac{l}{2l+3}|M_{l+1,l}|^2\right\}, \nonumber \\
	& \frac{d\sigma}{d\omega} = \frac{\alpha}{2\pi^3\omega}\frac{p_{f}}{p_{i}}\,a\,.
\end{align}

\end{document}